
\newif\iffigs\figstrue

%
\let\useblackboard=\iftrue
%
\let\useblackboard=\iffalse
%
%

\input harvmac.tex

\iffigs
  \input epsf
\else
  \message{No figures will be included.  See TeX file for more
information.}
\fi

%
%
%
\def\yboxit#1#2{\vbox{\hrule height #1 \hbox{\vrule width #1
\vbox{#2}\vrule width #1 }\hrule height #1 }}
\def\fillbox#1{\hbox to #1{\vbox to #1{\vfil}\hfil}}
\def\ybox{{\lower 1.3pt \yboxit{0.4pt}{\fillbox{8pt}}\hskip-0.2pt}}

\def\comments#1{}

\def\half{{1\over 2}}

\def\II{\relax{I\kern-.07em I}}

\def\IZ{\relax\ifmmode\mathchoice
{\hbox{\cmss Z\kern-.4em Z}}{\hbox{\cmss Z\kern-.4em Z}}
{\lower.9pt\hbox{\cmsss Z\kern-.4em Z}}
{\lower1.2pt\hbox{\cmsss Z\kern-.4em Z}}\else{\cmss Z\kern-.4em
Z}\fi}
\def\IB{\relax{\rm I\kern-.18em B}}

\def\ID{\relax{\rm I\kern-.18em D}}
\def\IE{\relax{\rm I\kern-.18em E}}
\def\IF{\relax{\rm I\kern-.18em F}}
\def\IG{\relax\hbox{$\inbar\kern-.3em{\rm G}$}}
\def\IGa{\relax\hbox{${\rm I}\kern-.18em\Gamma$}}
\def\IH{\relax{\rm I\kern-.18em H}}
\def\II{\relax{\rm I\kern-.18em I}}
\def\IK{\relax{\rm I\kern-.18em K}}
\def\IP{\relax{\rm I\kern-.18em P}}

\useblackboard
\def\IZ{\relax\Bbb{Z}}
\fi

\font\cmss=cmss10 \font\cmsss=cmss10 at 7pt
\def\IR{\relax{\rm I\kern-.18em R}}

\def\tilde{\widetilde}


\iffigs
  \input epsf
\else
  \message{No figures will be included.  See TeX file for more
information.}
\fi




\def\lim{{lim}}

\input epsf
















\def\hepth#1{{\it hep-th/{#1}}}

\def\frac#1#2{{{{#1}}\over {{#2}}}}           

\def\e8{E_8 \times E_8}                       
\def\HALFSONE{{\bf S}^{1}/{\bf Z}_2}    








%
\Title{ \vbox{\baselineskip12pt\hbox{hep-th/9803088, PUPT-1776}}}
{\vbox{
\centerline{Heterotic Matrix theory with Wilson lines}
\centerline{on the lightlike circle}}}
\centerline{Morten Krogh}
\smallskip
\smallskip
\centerline{Department of Physics, Jadwin Hall}
\centerline{Princeton University}
\centerline{Princeton, NJ 08544, USA}
\centerline{\tt krogh@princeton.edu}
\bigskip
\bigskip
\noindent

We develop a matrix model for the $SO(32)$ Heterotic 
string with certain Wilson lines on the lightlike circle. 
This is done by using appropriate T-dualities. 
The method works for an infinite number of Wilson 
lines, but not for all. 
The matrix model is the theory on the D-string 
of type I wrapped on a circle with a $SO(32)$ Wilson 
line on the circle. The radius of the circle depends on 
the Wilson line. This is a 1+1 dimensional $O(N)$ 
theory on a circle. N depends not only on the 
momentum around the lightlike circle but also 
on the winding and $SO(32)$ quantum numbers of the 
state. Perspectives for obtaining a matrix model 
for all Wilson lines are discussed.

\Date{March, 1998}



\lref\Connes{Alain Connes, Michael R. Douglas and Albert Schwarz,
  {``Noncommutative Geometry and Matrix Theory: Compactification on 
     Tori,''} \hepth{9711162}}

\lref\DougHull{Michael R. Douglas and Chris Hull,
  {\it ``D-branes and the Noncommutative Torus,''}
  \hepth{9711165}}

\lref\Narainet{K.S. Narain,
  {\it ``New Heterotic String Theories in Uncompactified 
         Dimensions <10'', Physics Letters 169B (1986) 41-46}}

\lref\Narain{K. Narain, M. Sarmadi and E. Witten,
    {\it ``A Note on Toroidal Compactification of Heterotic 
           String Theory'',
  Nucl. Phys. B279 (1987) 369.}}

\lref\Ginsparg{P. Ginsparg,
  Phys. Rev. D35 (1987) 648.}

\lref\Sen{Ashoke Sen,
  {\it ``D0 Branes on $T^n$ and Matrix Theory,''} \hepth{9709220}}

\lref\MotlBanks{Tom Banks and Lubos Motl,
  {\it ``Heterotic Strings from Matrices,''}
  \hepth{9703218}}

\lref\Horava{Petr Horava,
  {\it ``Matrix Theory and Heterotic Strings on Tori'',
    Nucl. Phys. B505 (1997) 84-108,} \hepth{9705055}}

\lref\Lowe{David A. Lowe,
  {\it ``Heterotic Matrix String Theory,
   Phys.Lett.B403 (1997) 243-249'',}
  \hepth{9704041}}

\lref\Kim{Nakwoo Kim and Soo-Jong Rey,
  {\it ``M(atrix) Theory on an Orbifold and Twisted Membrane,
   Nucl.Phys.B504(1997) 189-213,''}
    \hepth{9701139}}

\lref\Govin{Suresh Govindarajan,
  {\it ``Heterotic M(atrix) theory at generic points in Narain moduli space,
   Phys.Rev.D56(1997)5276-5278,''}
  \hepth{9707164}}

\lref\Seiwhy{N. Seiberg,
  {\it ``Why is the Matrix Model Correct?, 
   Phys. Rev. Lett. 79 (1997) 3577-3580,''} 
\hepth{9710009}}

\lref\Eva{Shamit Kachru and Eva Silverstein,
  {\it ``On Gauge Bosons in the Matrix Model Approach to M Theory,''
   Phys.Lett.B396(1997)70-76,}
  \hepth{9612162}}

\lref\Lubos{Lubos Motl and Leonard Susskind,
  {\it ``Finite N Heterotic Matrix Models and Discrete Light Cone
   Quantization,''}
  \hepth{9708083}}

\lref\Motlscrew{Lubos Motl,
  {\it ``Proposals on nonperturbative superstring interactions,''}
  \hepth{9701025}}

\lref\BS{Tom Banks and Nathan Seiberg,
  {\it ``Strings from Matrices,'' Nucl.Phys. B497 (1997) 41-55, }
  \hepth{9702187}}

\lref\Rey{Soo-Jong Rey,
  {\it ``Heterotic M(atrix) Strings and Their Interactions,'' 
   Nucl. Phys B502 (1997) 170-190,}
  \hepth{9704158}}

\lref\Kabat{Daniel Kabat and Soo-Jong Rey,
  {\it ``Wilson Lines and T-Duality in 
    Heterotic M(atrix) Theory,'' Nucl. Phys. B508 (1997) 535-568,}
  \hepth{9707099}}

\lref\Giveon{A. Giveon, M.Porrati and E. Rabinovici,
  {\it ``Target Space Duality in String Theory,'' Phys.Rept.244(1994)
     77-202}
  \hepth{9401139}}

\lref\Morten{Morten Krogh,
  {\it ``A Matrix Model for Heterotic $Spin(32)/Z_2$ and type I
    String Theory'',} 
  \hepth{9801034}}

\lref\Duiliu{Duiliu-Emanuel Diaconescu and Jaume Gomis,
  {\it ``Matrix Description of Heterotic Theory on K3'',} 
  \hepth{9711105}}

\lref\PeiWuWu{Pei-Ming Ho, Yi-Yen Wu and Yong-Shi Wu,
  {\it ``Towards a Noncommutative Geometric Approach to 
          Matrix Compactification'',} 
  \hepth{9712201}}

\lref\Peiadv{Pei-Ming Ho and Yong-Shi Wu,
  {\it ``Noncommutative Gauge Theories in Matrix 
         Theory'',} 
  \hepth{9801147}}

\lref\Casal{R. Casalbuoni,
  {\it ``Algebraic treatment of compactification on 
         noncommutative tori'',} 
  \hepth{9801170}}

\lref\MortenEdna{Yeuk-Kwan E. Cheung and Morten Krogh,
  {\it ``Noncommutative Geometry from 0-branes in a 
         Background B-field'',} 
  \hepth{9803031}}

\lref\Kawano{Teruhiko Kawano and Kazumi Okuyama,
  {\it ``Matrix Theory on Noncommutative Torus'',} 
  \hepth{9803044}}

\lref\Wati{Washington Taylor,
  {\it ``D-brane field theory on compact spaces'',
  Phys.Lett. B394 (1997) 283-287} \hepth{9611042}}

\lref\WOS{Ori J. Ganor, Sanjaye Ramgoolam and Washington Taylor,
  {\it ``Branes,Fluxes and Duality in M(atrix)-Theory'',
  Nucl.Phys.B492 (1997) 191-204} \hepth{9611202}}

\newsec{Introduction}

During the last year a matrix model for the $E_8 \times E_8$ 
Heterotic string theory has been 
developed \refs{\Eva, \Motlscrew, \Kim, \BS, \MotlBanks, \Lowe, \Rey, 
\Horava, \Kabat,
 \Govin, \Lubos}. The model can be derived by following Sen's and Seiberg's  
 prescription \refs{\Sen , \Seiwhy } for M-theory on $\HALFSONE$. The 
result is that Heterotic $E_8 \times E_8$ is 
described by the decoupled theory of D-strings in type I wound on a 
circle with a Wilson line on the circle breaking $SO(32)$ to $SO(16) \times 
SO(16)$. It is important to note that this matrix theory is a description 
of the $\e8$ Heterotic string with a Wilson line on the lightlike circle, 
which breaks $\e8$ down to $SO(16) \times SO(16)$. 

Recently a matrix model for the SO(32) Heterotic string was derived \Morten.
Here the theory had no Wilson line on the lightlike circle. So the situation, 
by now, is that two 10 dimensional Heterotic matrix models are known: The 
$\e8$ theory with a Wilson line on the lightlike circle breaking 
$\e8$ to $SO(16) \times SO(16)$ and the SO(32) theory with unbroken gauge 
group. Matrix models for Heterotic strings compactified on tori and K3 are 
also known \refs{\Horava, \Govin, \Duiliu}.
 In this paper we will only be interested in the 10 dimensional case.

The purpose of this paper is to find matrix models for other Wilson lines 
of the SO(32) theory. The motivation to search for such theories come 
from recent developments in M theory on $T^2$ \refs{\Connes, \DougHull, 
\PeiWuWu, \Peiadv, \Casal, \MortenEdna, \Kawano}, 
where a matrix model was derived in the case of a non-zero
3-form potential, $C_{-12} \neq 0$. The index $-$ denotes the lightlike 
direction and $1,2$ the $T^2$. The important feature is that 
background fields are turned on along the lightlike circle. The resulting 
matrix model is a SYM theory on a noncommutative torus. 
The advantage of noncommutative geometry is that it works for all 
values of $C_{-12}$ and that it is continous in $C_{-12}$. The drawback of 
the noncommutative geometry approach is that the theories are nonlocal 
and it is not clear whether they make sense as quantum theories.

However for a dense 
subset of values of $C_{-12}$ the resulting matrix model can be obtained 
using standard T-dualities without use of noncommutative geometry. One 
might argue that this is good enough for physical applications. The problem, 
however, is that the involved T-dualities depend crucially on the value of 
$C_{-12}$ giving a description of the physics which has no manifest 
continuity in $C_{-12}$.

It is natural to believe that the Heterotic theories behave similarly. This is 
especially so for the SO(32) theory, where the description without a 
Wilson line is known \Morten. In this paper we will show that for a 
countable set of Wilson lines of the SO(32) Heterotic string theory 
standard T-dualities will provide us with a matrix description of 
the theory. Furthermore we will briefly speculate on perspectives 
for obtaining a noncommutative model for the Heterotic string.

The organization of the paper is as follows. In section 2 we 
briefly discuss the case of M theory on $T^2$ because of its similarity 
with the Heterotic case. In sction 3 we discuss T-dualitites of the 
SO(32) Heterotic string and show how to obtain a matrix model for a 
countably infinite number of Wilson lines. In section 4 we discuss perspectives 
for obtaining a noncommutative model for all Wilson lines. We end with 
a conclusion in section 5.


\newsec{M theory on $T^2$}

Let us first recall how to obtain the matrix model for M theory on $T^2$ 
following {\Sen, \Seiwhy}. For the moment we set $C_{-12}=0$. Let the 
lightlike circle have radius R, the Planck mass be M and momentum 
$N \over R$. The radii of the $T^2$ are called $R_1,R_2$. For simplicity 
we take the torus to be rectangular. By Seiberg's boost and rescaling 
\Seiwhy\ we are led to consider a theory on a spatial circle of 
radius $\tilde R$, Planck mass $\tilde M$ and radii $\tilde R_1,\tilde R_2$ 
in the limit $\tilde R \rightarrow 0$ with
\eqn\fikse{\eqalign{ 
{\tilde M}^2 {\tilde R} &= M^2 R \cr
{\tilde M} {\tilde R_i} &= M R_{i} \qquad i=1,2 
}}
This turns into N D0-branes in type IIA with string 
mass $m_s$ and coupling $\lambda$ 
on a $T^2$ with radii $\tilde R_i$ where
\eqn\toa{\eqalign{
m_s^{2} &= {\tilde M}^3 {\tilde R} \cr
\lambda  &= ({\tilde M \tilde R})^{3/2}
}}
Now the transverse radii $\tilde R_i$ shrink so we perform a T-duality 
on both circles. This finally gives a theory with N D2-branes on a $T^2$ 
with radii
\eqn\radier{
R_{i}^{'} = {1 \over m_s^2 \tilde R_i} = {1 \over M^3 R R_i}
}
which are finite. This is a U(N) gauge theory. The crucial step was the 
T-duality on the two transverse circles. A T-duality turns a circle 
with radius going to zero into a circle with a radius going to 
infinity. This is so for fixed string mass. In our case the string mass 
goes to infinity so the final circle actually has finite size. 

Let us now turn on a background field $C_{-12} \neq 0$. Following 
the same procedure as before we get to type IIA with a B field, $B_{12}
\neq 0$. The radii of the torus are shrinking exactly as before. We 
would like to perform a T-duality which makes the radii large. The
standard T-duality, used above, does not do the job. When $B_{12} \neq 
0$ it relates a shrinking torus to another shrinking torus as we 
explain now.

First we must recall how T-duality works for 
type IIA on a $T^2$. The compactification is, among others,
 specified by 4 real parameters. These are the complex structure, 
$\tau$, of the torus, the volume, $V$, and the integral of the B-field 
over the torus, $B$. $V$ and $B$ are conveniently collected into 
a complex number in the upper halfplane, $\rho = B + iV$. Here 
$V$ is measured in string units, so $\rho$ is dimensionless. The 
T-duality group \Giveon\ is now $SL(2,Z) \times SL(2,Z)$ where 
the first factor acts on $\tau$ and the second on $\rho$. There
 is also a T-duality which exchanges $\tau$ and $\rho$, but that 
will take us from type IIA to type IIB. We will only be concerned 
with $\rho$ and the associated $SL(2,Z)$. $V$ goes to zero in 
Seiberg's limit, so we would like to make a T-duality that takes
 us to large $V$. In the following we will work at 
fixed string mass. When needed the string mass can easily be 
reinstated.

Let us perform the T-duality given by
\eqn\slmat{
\pmatrix{a & b \cr c & d \cr} \in SL(2,Z)
}
The transformed $\rho$ is
\eqn\transformed{
\rho^{'} = { a\rho + b \over c\rho +d}  
 = { a(B +iV)+b \over c(B+iV)+d}
} 
giving
\eqn\bv{\eqalign{
B^{'} &= {(aB+b)(cB+d)+ acV^2 \over (cB+d)^2 + c^2 V^2} \cr 
V^{'} &= {V \over  (cB+d)^2 + c^2 V^2}
}}
We see that $V^{'} \rightarrow 0$ unless $cB+d =0$ in which 
case $V^{'} \rightarrow \infty$. We conclude that exactly for 
rational values of $B$ can we obtain a matrix model in this way. 
Let us look more closely at what model we obtain in the 
case $cB+d=0$. Suppose 
we start in a sector with N D0-branes and $N_2$ membranes 
wrapped around the $T^2$. After the T-duality we have
\eqn\nulto{
\pmatrix{N^{'} \cr  -N_2^{'}} = \pmatrix{a & b \cr c & d \cr}
\pmatrix{N \cr  -N_2}
}
Especially $N_2^{'}= -cN +dN_2$ so we do not get a U(N) theory but 
a $U(| -cN+dN_2 |)$ theory. The volume of the torus is $V^{'}= 
{1 \over c^2V}$
which is reduced by a factor of $1 \over c^2$ compared to the case 
$B=0$. Furthermore there is a background field $B^{'}={a \over c}$. 
We see that for various rational values of $B = -{d \over c}$ we get 
gauge theories with different unitary groups on tori of different 
size. The achievement of noncommutative geometry is that it can give a 
continous description of the physics.

Before we go on to the main interest of this paper, namely the 
Heterotic string, let us understand in another way why any rational 
value of B can be coped with by T-dualitites. This will prove useful 
for understanding the Heterotic T-dualities in the next section. 
The group $SL(2,Z)$ is generated by the two elements S and T.
\eqn\st{
S= \pmatrix{0 & -1 \cr 1 & 0} \qquad T = \pmatrix{1 & 1 \cr 0 & 1}
}
S is the usual volume inversion when $B=0$ and T shifts the 
B field by one. We are considering $\rho = B +iV$ in the limit 
$V \rightarrow 
0$. We now observe that setting $V=0$ and then acting with S or T or 
acting with S or T and then setting V=0 amounts to the same thing 
with one exception. This exception is acting with S in the case $B=0$. 
This observation follows easily from the explicit expressions
\eqn\explic{\eqalign{
T: \qquad & B+iV \rightarrow B+1+iV \cr
S: \qquad & B+iV \rightarrow {-B \over B^2 + V^2} + i{V \over B^2 +V^2}
}}
This means that we can use the following strategy given a B. We put 
$V=0$ and act with a product of powers of S and T until B=0.
Finally acting with S we get to infinite volume as we want. Let us 
illustrate this with an example. Suppose $B={2 \over 3}$. We then 
have the following group actions.
\eqn\groupact{
B={2\over3} \buildrel T^{-1} \over \longrightarrow
B= -{1\over3} \buildrel S \over \longrightarrow
B= 3 \buildrel T^{-3} \over \longrightarrow
B=0
}
In other words the matrix 
\eqn\doesthejob{
ST^{-3}ST^{-1} = \pmatrix{-1 & 1 \cr -3 & 2}}
does the job. We did not learn much new from this since we solved 
the problem above. However the equivalent method will prove useful 
in the Heterotic case, where the T-duality group is more complicated.


\newsec{SO(32) Heterotic String theory}

In this section we will first review various well known facts about 
T-duality in Heterotic theories \refs{\Narainet, \Narain}. We will 
then use this information to show that a matrix model for the SO(32) 
Heterotic string can be obtained for a countable number of Wilson 
lines on the lightlike circle. This extends the result in \Morten\ 
where the matrix model was derived in the case of no Wilson line.

\subsec{Heterotic T-dualities}
The moduli space for Heterotic string theory compactified on a 
circle is\refs{\Narainet, \Narain} 
\eqn\modulirum{
O(17,1,Z) \backslash O(17,1,{\bf R})/O(17) \times O(1).
}
There is no distinction between 
$\e8$ and SO(32) as soon as we compactify on a circle. 
We will think of this moduli space in SO(32) terms, since this is 
the case of relevance for us. Let us try to understand the 
moduli space more concretely. Consider the space ${\bf R}^{17,1}$ 
with metric $g_{\mu \nu}= diag(1,\ldots,1,-1)$. In this space we 
place the even selfdual lattice 
$\Gamma^{17,1}= \Gamma^{16} + \Gamma^{1,1}$. $\Gamma^{16}$ is the 
weightlattice of $spin(32)/Z_2$ which can be described as 
the set of points $(a_1,\dots,a_{16})$ where either all 
$a_i \in {\bf Z}$ or all $a_i \in {\bf Z}+ \half$ and in both 
cases $\sum_i a_i$ is even. $\Gamma^{1,1}$ is the lattice 
consisting of points ${1 \over \sqrt{2}}(m+n,m-n)$ for $m,n \in 
{\bf Z}$. Consider the basis vectors 
$e_i= (0,\ldots,0,1,0,\ldots,0)$ of ${\bf R}^{17,1}$. The 
first 17 of these, $e_i \;\;\;
i=1,\ldots,17$ span a 17 dimensional subspace of positive signature 
and $e_{18}$ span a 1 dimensional subspace of negative signature. 
Let $g \in O(17,1,{\bf R})$, then $g e_i$, $i=1,\ldots,17$ span a  
17 dimensional subspace of positive signature and $g e_{18}$ span 
an orthogonal 1 dimensional subspace of negative signature. Every such 
subspaces of ${\bf R}^{17,1}$ can be obtained for appropriate $g$. 
If $h \in O(17) \times (1)$, then $gh$ obviously gives the same 
subspaces as $g$; the basis have just changed. This gives an 
identification of $O(17,1,{\bf R})/O(17) \times O(1)$ with 
the space of 17 dimensional subspaces of positive signature. 
Note that the 1 dimensional subspace is determined as the 
orthogonal complement of the 17 dimensional space. 

Now let $P \in \Gamma^{17,1}$. $P$ represents, together with 
other quantum numbers, a state in the theory. Let also a 
$g \in O(17,1,{\bf R})/O(17) \times O(1)$ be given. $g$ determines 
two subspaces as explained above. Decompose $P$ as $P_L + P_R$,
 where $P_L$ is in the 17 dimensional subspace and $P_R$ is in 
the 1 dimensional subspace. This is a unique decomposition. In 
terms of the basis $e_i$,
\eqn\dekomp{\eqalign{
&(P_L)_i = < ge_i , P> \qquad i=1,\dots,17 \cr
&(P_R) = -<ge_{18} , P> 
}}
where $<,>$ denotes the inner product in ${\bf R}^{17,1}$. 

The space $O(17,1,{\bf R})/O(17) \times O(1)$ is 17 dimensional. 
It is not hard to see that the space can be parametrized by 
a vector $A \in {\bf R}^{16}$ and a number $R>0$. As explained 
in \Narain\ $R$ can be understood as the radius of the circle and 
$A$ is the Wilson line. Given a lattice point specified by $Q \in 
 \Gamma^{16}$ and the integers $m,n$ labeling $\Gamma^{1,1}$ the 
left- and right-moving momenta are \Ginsparg
\eqn\PRA{\eqalign{
P_L &= ( Q + An, {1 \over \sqrt 2} ({ m - \half A^2n - AQ 
\over R}+ nR)) \cr
P_R &= {1 \over \sqrt 2} ({ m - \half A^2n - AQ 
\over R}- nR)
}}

Let us now consider the T-duality group $O(17,1,Z)$. This is 
the group of $O(17,1)$ matrices which preserve the lattice 
$\Gamma^{17,1}$. The $Z$ does not mean that the entries in 
the matrix are integers, but that the lattice $\Gamma^{17,1}$ 
is preserved. A T-duality transformation by an element 
$t \in O(17,1,Z)$ means that we map the lattice into itself 
before we project into left- and rightmovers. Since the theory 
is determined by the full set of states and not by how we label 
them, this is a symmetry of the theory. Let $P \in \Gamma^{17,1}$. 
The T-duality transformation maps it into $tP$. The decomposition 
into left- and right-movers is
\eqn\tleftright{\eqalign{
(P_L)_i &= < ge_i, tP> = < t^{-1}ge_i,P> \cr
(P_R)_i &= < ge_{18},tP> = < t^{-1}ge_{18},P> 
}}
where we used that the adjoint of $t$ is $t^{-1}$ for 
$t \in O(17,1,Z)$. This shows that $g$ and $t^{-1}g$ define 
the same point in the moduli space.

Consider a point $P \in \Gamma^{17,1}$ and a $g \in O(17,1)$. 
$P$ is specified by $(Q,m,n)$. $g$ corresponds to $(R,A)$ as above.
Consider also the point $tP \in \Gamma^{17,1}$ specified by 
$(Q^{'},m^{'},n^{'})$ and $tg \in O(17,1)$ corresponding to 
$(R^{'},A^{'})$. Since $t$ is in $O(17,1,Z)$ the 
 decomposition of $P$ into $(P_L ,P_R)$ with respect to $g$ is 
the same as the decomposition of $tP$ into $(P^{'}_L,P^{'}_R)$ 
with respect to $tg$. We might thus think that if we plug into 
eq.\PRA\ with primed and unprimed variables we get the same 
result. This is the not true since the set $(R,A)$ corresponds 
to an element in $O(17,1)/O(17)\times O(1)$ and not an element 
in $O(17,1)$. In comparing $(P_L,P_R)$ with $(P^{'}_L,P^{'}_R)$
 obtained from eq.\PRA\ we should thus allow for for a $O(17)
 \times O(1)$ transformation. This might seem a bit 
complicated but we are saved by the fact that $O(1)$ is a very 
simple group. It consists of $\{ \pm 1 \}$. We thus conclude that 
for a T-duality given by $t \in O(17,1,Z)$
\eqn\mnQ{
{ m - \half A^2n - AQ \over R}- nR =
 \pm  ({ m^{'} - \half A^{'2}n^{'} - A^{'}Q^{'} 
\over R^{'}}- n^{'}R^{'})
}
The sign is actually easily determined from $t$. It is equal 
to the sign of the matrix entry $t_{18,18}$.

Formula \mnQ\ will be important to us. We will use it to determine 
$A^{'},R^{'}$ as a function of $A,R$ for a given T-duality. 
For a given T-duality, $(m^{'},n^{'},Q^{'})$
are known in terms of $(m,n,Q)$. The relation is a linear 
function. Collecting the coefficients of $(m,n,Q)$ in eq.\mnQ\ 
 will give equations which determine $(A^{'},R^{'})$ in terms 
of $A,R$.

Let us now go on to consider the structure of $O(17,1,Z)$.
We will list 3 different types of elements which we will 
use to obtain matrix models.
\item{$\underline{Case 1.}$}
\eqn\maten{
 T_U = \pmatrix{U_{16\times 16} & 0 \cr 
                0 & 1_{2 \times 2} \cr}
}
where $U \in O(16,Z)$. Here $Z$ means that $\Gamma^{16}$ is 
preserved. This T-duality transformation acts as follows
\eqn\Tet{\eqalign{
Q^{'} &= UQ \cr
m^{'} &= m \cr
n^{'} &= n 
}}
Solving eq.\mnQ\ we get
\eqn\RAet{\eqalign{
R^{'} &= R \cr
A^{'} &= (U^{t})^{-1} A
}}
\item{$\underline{Case 2.}$}
\eqn\matto{
 T_P = \pmatrix{1_{16\times 16} & P_i \over \sqrt 2 & 
        -P_i \over \sqrt 2  \cr 
        -P_i \over \sqrt 2 & 1- {1\over 4} P^2 & 
        {1 \over 4}P^2 \cr
         -P_i \over \sqrt 2 & - {1\over 4} P^2 & 
        1+ {1 \over 4}P^2 \cr}
}
with $P \in \Gamma_{16}$. This transformation acts as follows
\eqn\Tto{\eqalign{
Q^{'} &= Q + nP \cr
m^{'} &= m -\half n P^2 -QP \cr
n^{'} &= n 
}}
Solving eq.\mnQ\ we get
\eqn\RAto{\eqalign{
R^{'} &= R \cr
A^{'} &= A - P
}}
\item{$\underline{Case 3.}$}
\eqn\mattre{
 S = \pmatrix{1_{16\times 16} & 0 \cr 
                0 & {\pmatrix{-1 & 0 \cr 0 & 1 \cr}} \cr }
}
Here 
\eqn\Ttre{\eqalign{
Q^{'} &= Q \cr
m^{'} &= -n \cr
n^{'} &= -m 
}}
Solving eq.\mnQ\ we get
\eqn\RAtre{\eqalign{
R^{'} &= {R \over R^2 + \half A^2} \cr
A^{'} &= {A \over R^2 + \half A^2}
}}

\subsec{Matrix models}
Now we will consider the task of finding a matrix 
model for the SO(32) Heterotic string with a Wilson line on 
the lightlike circle. Let us first briefly review the 
method for obtaining this matrix model in the case of no 
Wilson line \Morten. We start with the SO(32) Heterotic 
theory with string mass M, coupling $\lambda$ and radius 
of the lightlike circle R. We consider the sector with momentum 
$N \over R$. First we follow Seiberg \Seiwhy\ and relate it to 
a spatial compactification on a shrinking circle. We now perform 
the T-duality called $S$ above. In the case $A=0$ it sends 
$R \rightarrow {1 \over R}$ and exchanges momentum and 
winding. This takes us to a sector with winding N on a large 
circle. Finally we employ the Heterotic SO(32)-type I 
duality. This produces a theory on N D-strings of type I 
wrapped on a circle. This is an O(N) gauge theory.

If $A \neq 0$ this chain of dualities will not work, 
since the T-duality transformation, $S$, only maps a small 
circle into a large circle when $A=0$. However by analogy 
with M theory on $T^2$ we could try another T-duality 
transformation. If we can find a T-duality transformation 
that makes the circle large we will have obtained a matrix model, 
since the type I- SO(32) Heterotic duality works in any 
case. The type I coupling does go to zero and the theory on the 
D-strings decouples.

Let us now show that for a countable number of Wilson lines 
such a T-duality transformation actually exists. We will 
follow a procedure similar to the one described in the end 
of section 2 for M theory on $T^2$.

First we notice that the limit $R \rightarrow 0$ commutes with 
the T-duality elements $T_U,T_P$ and $S$ except for $S$ when 
$A=0$. Given a Wilson line $A$ we should find a product of 
$T_U,T_P$ and $S$ which takes $A$ to zero. All this should be 
done for $R=0$. The resulting T-duality transformation 
succeeded by $S$ will then do the job. We could also reverse 
the process and start with any product of $T_U,T_P$ and $S$, 
not ending with $S$. The inverse of that will then map $A=0$ 
into some $A$. This A will then be an A for which a matrix 
model exist. We see that this exactly produces a countable 
set of Wilson lines for which a matrix model can be obtained 
using this approach. They certainly all have rational entries. 
Unlike the case of M theory on $T^2$ it is not all Wilson 
lines with rational entries which can be solved in this way.

Let us take an explicit example. Start with Wilson line
$A=(\half,0,\dots,0)$ and a shrinking spatial circle, $R_s 
\rightarrow 0$. This Wilson line breaks the gauge group 
down to $U(1) \times SO(30)$. Performing the duality 
$ST_PS$ with $P=(4,0,\ldots,0)$ we easily calculate that 
the resulting radius and Wilson line are 
$R_s^{'} = {1 \over 8R_s}$, $A^{'}=-A$. After dualizing to 
type I it becomes an $O(N^{'})$ theory on a circle of 
radius $1\over 8$ of what was the case for $A=0$ and 
with a Wilson line $A^{'}$ around the circle. The number
$N^{'}$ in $O(N^{'})$ is 
\eqn\gruppe{
N^{'}= -(n + 8m - Q \cdot (4,0,\ldots,0))
}
We see that it is not just equal to the momentum m.

Now we have seen, that lots of Wilson lines can be 
coped with in this way, it is natural to ask whether 
the $SO(16)\times SO(16)$ point is one of these. This 
would be strange since the resulting matrix model would 
be the theory on D-strings of type I wound on a circle 
with a Wilson line breaking SO(32) to $SO(16) \times 
SO(16)$. This model is known to be the matrix model for 
$\e8$ Heterotic string theory on the  $SO(16) \times 
SO(16)$ point. It would lead to a contradiction if this 
model was also a matrix model for the SO(32) theory.
We will now argue that, in fact, the  $SO(16) \times 
SO(16)$ Wilson line cannot be coped with using the 
method described above. In this method a chain of 
T-dualities is employed which maps a set $(R_s,A)$ into 
a set $(R_s^{'},A^{'})$ such that $A^{'} \rightarrow 0$ 
in the limit $R_s \rightarrow 0$. This T-duality is then 
succeeded by $S$ which maps to a large circle. However 
T-duality also preserve the gauge group and there does 
not exist any Wilson lines close to zero which breaks 
SO(32) to $SO(16) \times SO(16)$. The Wilson lines doing
 that form a discrete set. The same argument could be used 
to rule out many other unbroken gauge groups.


\newsec{Perspectives for obtaining a matrix model for all 
        Wilson lines}

So far we have seen that it is possible to obtain a matrix model 
for a countable number of Wilson lines. What about the general case? 
Unfortunately we do not have such a model. It would presumably be 
similar in spirit to the models for M-theory on $T^2$ with a 
background three-form potential \Connes. This model was 
obtained by generalising the procedure developed in 
\refs{\Wati, \WOS}.

What could be the equivalent for $SO(32)$ Heterotic matrix 
theory. Without a Wilson line the model is the theory living 
on the D-string of type I wrapped on a circle. This model 
is equivalent to 
zero-branes in type IIA on a $S^1$ modded out by a $Z_2$ group
$\{ 1, \Omega R_9 \} $, where $\Omega$ is worldsheet orientation 
reversal and $R_9$ is reflection in the 9'th direction. We can 
view the $S^1$ as ${\bf R}/Z$ where $Z$ is a group of translations. 
This is the same as 0-branes in type IIA modded out by the 
group generated by the $Z$ and the $Z_2$ above. Doing it in the usual 
way would just give us the theory on the D-string of type I 
on a dual circle. However in the process one needs to introduce 
a ``twisted'' sector, namely the strings going from 0-branes
to 8-branes. Finding the most general way of modding out 
with this group, allowing extra noncommutativity 
in some way, might give us a candidate for Heterotic matrix 
theory with a Wilson line on the lightlike circle. 
Methods for modding out 0-brane mechanics have been considered 
in \refs{\PeiWuWu, \Peiadv}. What makes the situation harder 
in this case is that the noncommutativity is in the twisted 
sector, so one needs an improvement of the methods 
considered there. 

Unfortunately there does not seem to be an equivalent of the 
methods employed in \refs{\DougHull, \MortenEdna, \Kawano} for 
the Heterotic case.


\newsec{Conclusions}

Using T-dualities we have obtained a matrix model for the 
$SO(32)$ Heterotic string theory with certain Wilson lines 
on the lightlike circle. We do not know a method to obtain a 
model which works for all Wilson lines.

What about the $\e8$ theory? The matrix model in this theory 
is for the Wilson line breaking $\e8$ down to $SO(16) \times 
SO(16)$. It would be nice to have a larger unbroken gauge group, 
especially $\e8$. The methods described here can not be used to 
obtain that. The reason is that the matrix model for the $\e8$
theory is obtained by T-dualizing to an $SO(32)$ theory on a large 
circle. Since the unbroken gauge group is the same and there are 
no enhanced gauge symmetries for very small or very large circles 
the unbroken gauge group has to be a subgroup of both 
$\e8$ and $SO(32)$ which means that $SO(16) \times SO(16)$ is the 
best one can do using these methods.

It will be an interesting challenge for the future to find matrix 
models for all Wilson lines for both the $SO(32)$ and $\e8$ 
theories.

\bigbreak\bigskip\bigskip
\centerline{\bf Acknowledgments}\nobreak
I wish to thank Yeuk-Kwan E. Cheung and Ori Ganor
 for fruitful discussions. 
This work was supported by The Danish Research Academy.


\listrefs
\bye
\end